\documentclass[twocolumn]{webofc}
\usepackage[varg]{txfonts}
\usepackage{graphicx}
\usepackage{float}
\usepackage[usenames,dvipsnames]{color}
\usepackage{soul} 
\usepackage{amsmath} 
\begin{document}
\title{Reactions Governing Strangeness Abundance in Primordial Universe}
\author{\firstname{Cheng Tao} \lastname{Yang}\inst{1}
 \and
 \firstname{Johann} \lastname{Rafelski}\inst{1}
}

\institute{Department of Physics, The University of Arizona, Tucson, Arizona 85721, USA}
\date{\today}


\abstract{Strangness production processes can balance natural strangeness decay in the early hadronic Universe. Comparing to the characteristic Hubble time $1/H$, the reaction rates for $\mu^\pm+\nu_{\mu}\rightarrow K^\pm$, $l^-+l^+\rightarrow\phi$, and $\pi+\pi\rightarrow K$ in sequence become slower than expansion rate at $T=33.9\,\mathrm{MeV}$, $T=25\,\mathrm{MeV}$ and $T=20\,\mathrm{MeV}$ respectively. This means that in the antibaryon annihilation epoch near to $T\simeq 40$\,MeV strangeness is in chemical equilibrium. }
\maketitle

\noindent\textbf{Overview:} The present work is a first consideration of strangeness quark flavor evolution in the Universe following on the hadronization of quark-gluon plasma (QGP) near to $T_h=150$\,MeV. We characterize Universe expansion dynamics during the epoch of interest $T_h\approx 150\ge T\ge 10$\,MeV and compare strangeness population influencing reactions with the characteristic Hubble time $1/H$. 

We evaluate the dynamic rates employing detailed balance: the natural decay of particles concerned provides also the intrinsic strength of the inverse production reaction of interest. We outline consequences of our findings, and delimit related future research. Considering that baryons are a tiny component of all hadrons, this study is focused on the meson sector of the hadronic Universe.
 
We study reactions in which the particle number changes by one: An example is a particle decay  into two, restored in two to one reactions in what one may call one$\Leftrightarrow$two reaction process. For these the required reaction rates are obtainable from available decay rates as described below. The two$\Leftrightarrow$two reactions often have a significantly higher reaction threshold when strangeness pair abundance is produced. Threfore these reactions are important near to the QGP hadronization temperature $T_h\simeq 150$\,MeV. For discussion of the here relevant reactions, such as $KK\leftrightarrow\pi\pi$, we refer the reader to Chapter 18 in Ref.\cite{Letessier:2002gp}.\\[-0.2cm] 

\noindent\textbf{Universe expansion:} Since all reaction rates we obtain are functions of $T$ we also seek $1/H$ as a function of $T$. The Hubble parameter can be written in terms of energy density $\rho_\mathrm{tot}$~\cite{Kolb:1990vq}; $H^2=\frac{8\pi G_\mathrm{N}}{3}\,\rho_\mathrm{tot}(T)$, $G_\mathrm{N}$ is the gravitational constant. For $ 150\ge T\ge 10$\,MeV the Universe is radiation-dominated $\rho_\mathrm{tot}(T)\propto T^4$, thus we  easily express $1/H$  as a function of temperature  required as reference in our work. 

We include in $\rho_\mathrm{tot}$ photons, three flavors of neutrinos, charged leptons and antileptons $e^\pm,\mu^\pm$ and the lightest hadrons $\pi^0, \pi^\pm$. Both muons and pions are not relativistic in the considered $ 150\ge T\ge 10$\,MeV range. Evaluating their contribution to $\rho_\mathrm{tot}$ we use their actual mass and chemical equilibrium abundance: Muons are coupled through electromagnetic reactions $\mu^++\mu^-\Leftrightarrow\gamma+\gamma$ to the photon background and retain their chemical equilibrium~\cite{Rafelski:2021aey}. Because $\pi$-mesons are in thermal equilibrium~\cite{Kuznetsova:2008jt}, and are the lightest hadrons, they dominate hadronic contribution to the energy density near $T\approx 50$\,MeV.\\[-0.2cm]

\noindent\textbf{Strangeness in an expanding Universe:} Regarding initial conditions: We assume that before hadronization in the QGP epoch of temperature $T>150$\,MeV, the strangeness formation processes are fast enough to assure chemical equilibrium -- hence the Universe undergoing a relatively slow on hadronic time scale phase transformation emerges from QGP near to chemical equilibrium abundance. In this transition the excess QGP entropy, as compared to hadron gas (HG), is absorbed in additional Universe comoving volume expansion while excess strangeness has time to reequilibrate into equilibrium HG abundance. We note that:
\begin{itemize}
\item
The strange quark abundance in the early Universe is equal to the sum of strange baryons $ \Lambda^0,\,\Sigma^0,\,\Sigma^\pm$ and strange mesons $\phi,\,K,\,K^\pm$. We will see that in the early Universe the important sources for creation of $\phi(s\bar s)$ and $K(q\bar s)\overline{K}(\bar q s)$ are given by $l^-+l^+\rightarrow\phi$, $\rho+\pi\rightarrow\phi$, $\pi+\pi\rightarrow K,\overline{K}$, and $\mu^\pm+\nu\rightarrow K^\pm$.
\item
Reaction rate for $l^-+l^+\rightarrow\phi$, $\pi+\pi\rightarrow K$, and $\mu^\pm+\nu\rightarrow K^\pm$ are relatively small compared to reaction $\rho+\pi\rightarrow\phi$. However, the thermal abundance of pions and leptons $\pi$, $l^\pm$, and  $\nu$ is relatively high compared to the more massive $\rho$. Therefore all these reactions compete with each other.
\item
The strong decay rate of $\phi\rightarrow K+K$ indicates that the particles $\phi$ and $K$ are in relative chemical equilibrium.
\item
The strong strangeness exchange rate $K+N\rightarrow Y+\pi$ following on $\phi\rightarrow K+K$ with the small decay rates of $K\rightarrow\pi+\pi$ and $Y\rightarrow N+\pi$ suggest the possibility that strangeness can accumulate in $K$ meson and in $Y$ hyperons resulting in strange baryon yields above chemical equilibrium yield, a topic of future study requiring consideration of baryon chemical potential considered negligible in present study.
\end{itemize}

\begin{figure}[t]
\begin{center}
\includegraphics[width=0.95\columnwidth]{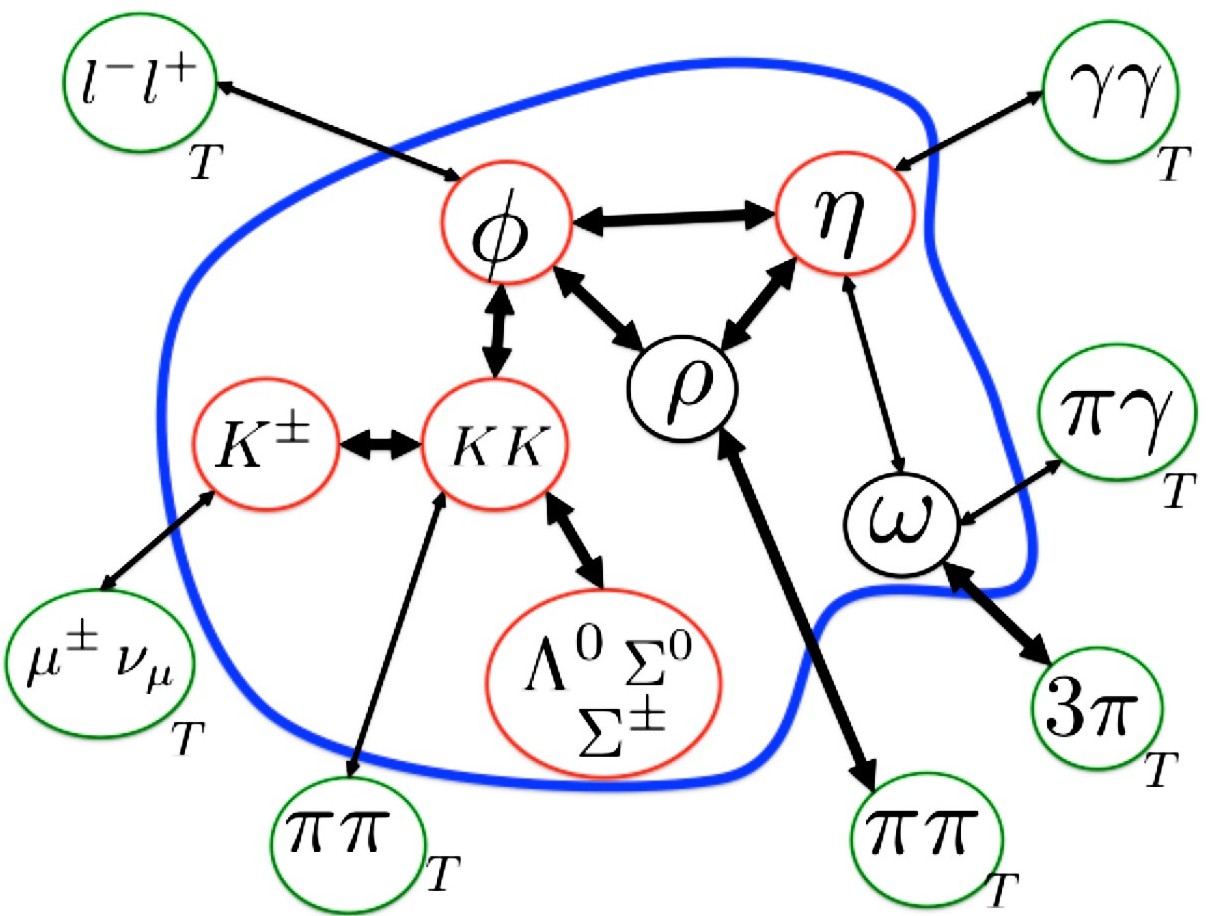}
\caption{Principal strangeness abundance changing processes in the hadronic Universe $T<T_h=150\,\mathrm{MeV}$. Within the (blue) boundary we see hadronic particles expected to fall out of abundance (chemical) equilibrium. The red circles within this domain represent strangeness-carrying mesons, black non-strange mesons of importance in creation of strangeness. Particles in green circles indicating their full equilibrium (subscript $T$) are seen outside this (blue) boundary domain.}
\label{Strangeness_map}
\end{center}
\end{figure}

The key meson production and decay processes are illustrated in Fig.~\ref{Strangeness_map} including important \lq hidden\rq\ $s\bar s$ components in $\eta$ and $\phi$, and multi-step reactions leading to formation of $\phi$. All particles outside the (blue) boundary are in abundance (chemical) and kinetic (thermal) equilibrium in the temperature domain of interest, see Ref.\,\cite{Kuznetsova:2008jt,Birrell:2014uka}.\\[-0.2cm]

\noindent\textbf{Strength of available reaction rates:} 
We introduce here the typical interaction rates relevant for the discussion of strangeness in the early Universe using the particle data group data tables~\cite{Tanabashi:2018oca}. We have
\begin{itemize}
 \item For $\phi$ meson, the partial decay rates are 
 \begin{align}
 &\Gamma_{\phi\rightarrow KK}=3.545\,\mathrm{MeV},\\
 &\Gamma_{\phi\rightarrow\rho\pi}=0.6535\,\mathrm{MeV},\\
 &\Gamma_{\phi\rightarrow\eta\gamma}=0.0558\,\mathrm{MeV},\\
 &\Gamma_{\phi\rightarrow ll}=2.484\times10^{-3}\,\mathrm{MeV}.
 \end{align}
 \item For $\eta$ meson, the partial decay rates are given by
 \begin{align}
&\Gamma_{\eta\rightarrow\gamma\gamma}=0.355\times10^{-3}\,\mathrm{MeV},\\
&\Gamma_{\eta\rightarrow3\pi^0}=0.516\times10^{-3}\,\mathrm{MeV},\\
&\Gamma_{\eta\rightarrow3\pi}=0.368\times10^{-3}\,\mathrm{MeV},\\
&\Gamma_{\omega\rightarrow\eta\gamma}=3.905\times10^{-3}\,\mathrm{MeV},\\
&\Gamma_{\rho\rightarrow\eta\gamma}=44.73\times10^{-3}\,\mathrm{MeV}.
 \end{align}
 \item For $\rho$ meson, the decay rate is 
 \begin{align}
 \Gamma_{\rho\rightarrow\pi\pi}=149.1\,\mathrm{MeV}.
 \end{align}
 \item For $K$ meson, the decay rate for $K_\mathrm{S}\rightarrow\pi+\pi$ and charge pion decay $K^\pm\rightarrow\mu^\pm\nu$ are given by
 \begin{align}
 &\Gamma_{K_\mathrm{S}\rightarrow\pi\pi}=7.350\times10^{-12}\,\mathrm{MeV},\\
 &\Gamma_{K^\pm\rightarrow\mu\nu}=3.379\times10^{-14}\mathrm{MeV}.
 \end{align}
 \item For hyperons $Y$, the decay rate is comparable to $\Gamma_{K_\mathrm{S}\rightarrow\pi\pi}$. We postpone discussion pending more complete study of (anti)baryon component and note here
 \begin{align}
 \Gamma_{Y\rightarrow N\pi}\approx6.582\times10^{-12}\,\mathrm{MeV}.
 \end{align}
\end{itemize}

\begin{figure}[ht]
\begin{center}
\includegraphics[width=0.9\linewidth]{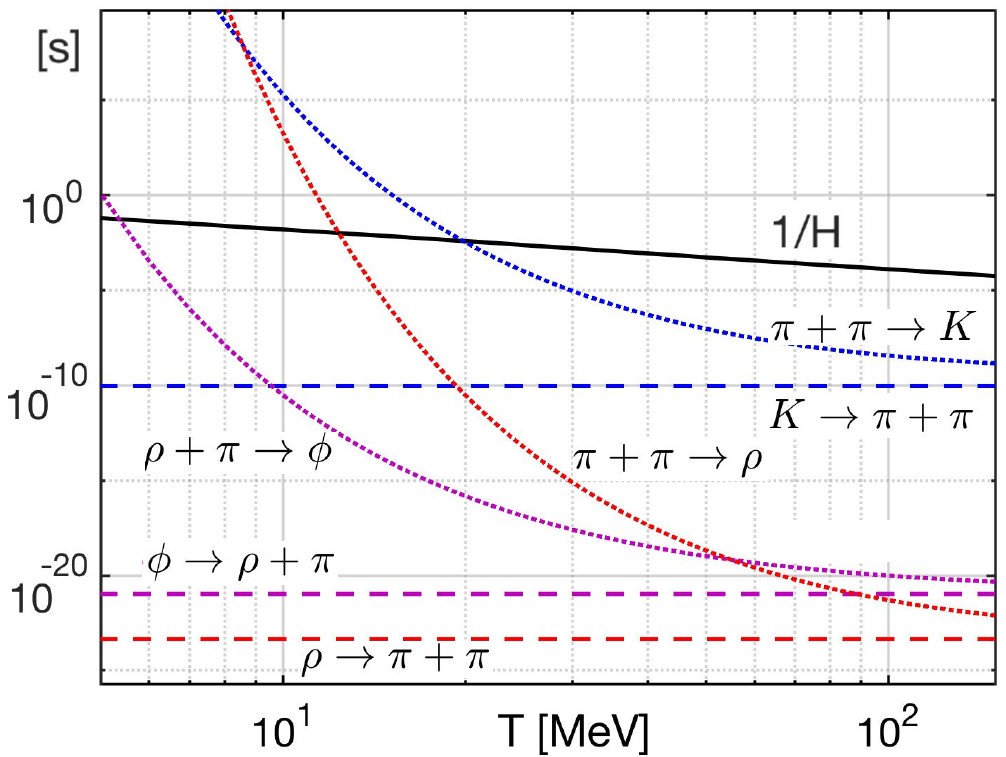}
\caption{Hubble time $1/H$ (black line) as a function of temperature is compared to hadronic relaxation reaction times, see Eq.\,(\ref{Reaction_Time}), for reactions $\pi+\pi\leftrightarrow K$(blue), $\pi+\pi\leftrightarrow\rho$ (red), $\rho\pi\leftrightarrow\phi$(purple). The horizontal dashed lines are the corresponding natural decay lifespans.}
\label{reaction_time_tot}
\end{center}
\end{figure}

\noindent\textbf{Hadron based strangeness production:} We now explore in more detail the conditions in which the nonequilibrium of strangeness arises. In order to study the strangeness nonequilibrium in the early Universe, we need to calculate the relevant dynamic reaction rates for strangeness production in detail. In the hadron sector the relevant interaction rates competing with Hubble time involving strongly interacting mesons are the reactions $\pi+\pi\leftrightarrow K$, $\pi+\pi\leftrightarrow\rho$, $\rho+\pi\leftrightarrow\phi$, and $\mu^\pm+\nu\rightarrow K^\pm$. Our results are seen in Fig.~\ref{reaction_time_tot}. We now turn to describe the methods used to obtain these results.

In general, the thermal reaction rate per time and volume for reaction $1+2\rightarrow3$ is given by \cite{Letessier:2002gp,Koch:1986ud}:
\begin{align}
R_{12\rightarrow3}&=n_1\,n_2\,\langle\sigma v\rangle_{12\rightarrow3}=\frac{g_1g_2}{32\pi^4}\,\frac{T}{1+I_{12}} \ \times\ \\
&\hspace*{-0.5cm}\int^\infty_{m^2_3}\!\!\!\sigma\,\frac{\big[s-(m_1+m_2)^2\big]\big[s-(m_1-m_2)^2\big]}{\sqrt{s}}K_1(\!\sqrt{s}/T)\,ds\notag\;,
\end{align}
where the factor $1/(1+I_{12})$ is introduced to avoid double counting of indistinguishable particles; we have $I_{12}=1$ for identical particles, otherwise $I_{12}=0$.
Many studies have shown that the hadronic reaction matrix element $|\mathcal{M}_{12\rightarrow 3}|^2$ for hadronic reactions without a resonance phenomenon is insensitive to changes of $\sqrt{s}$ in the considered energy range \cite{Letessier:2002gp,Koch:1986ud}. In this case, we assume that the transition amplitude $|\mathcal{M}_{12\rightarrow 3}|^2$ is nearly constant in the energy range we consider, and the cross section $\sigma$ can be written as
\begin{align}
\sigma=&\frac{2\pi/(g_1g_2)\sum_\mathrm{spin}|\mathcal{M}_{12\rightarrow3}|^2}{2\sqrt{\big[s-(m_1+m_2)^2\big]\big[s-(m_1-m_2)^2\big]}} \ \times \ \notag\\&\ \int d^4p_3\delta_0(p^2_3-m^2_3)\delta^4(p_1+p_2-p_3)\;.
\end{align}
In this case, the thermal reaction rate per time and volume can be written as
\begin{align}
R_{12\rightarrow 3}&=\frac{1}{16\pi^3}\,\frac{T}{1+I}\,\sum_{spin}|\mathcal{M}_{12\rightarrow 3}|^2\ \times\ \\&\frac{\sqrt{\big[m_3^2-(m_1+m_2)^2\big]\big[m^2_3-(m_1-m_2)^2\big]}}{2m_3}\,K_1(m_3/T)\notag\;.
\end{align}
On the other hand, the vacuum decay rate for particle $3$ can be written as
\begin{align}
&\frac{1}{\tau^0_3}=\frac{|p|}{8\pi\,m^2_3}\left(\frac{1}{1+I_{12}}\right)\left(\frac{1}{g_3}\sum_{spin}|\mathcal{M}_{12\rightarrow 3}|^2\right),\\
&|p|=\frac{1}{2m_3}\!\sqrt{m_1^4\!+m_2^4\!+m_3^4\!-2m^2_1m^2_2\!-2m^2_1m^2_3\!-2m^2_2m^2_3}\;.
\end{align} 

Hence, using the vacuum decay rate for the thermal reaction rate per unit time and volume can be written as
\begin{align}
\label{Thermal_Rate}
R_{12\rightarrow3}&=\frac{g_3}{2\pi^2}\left(\frac{T^3}{\tau^0_3}\right)\left(\frac{m_3}{T}\right)^2\,K_1(m_3/T)\;.
\end{align}
In order to compare the reaction time with Hubble time $1/H$, it is also convenient to define the relaxation time for the process $1+2\rightarrow 3$ as follow
\begin{align}
\label{Reaction_Time}
&\tau_{12\rightarrow3}\equiv\frac{dn_{1}/d\Upsilon_{1}}{R_{12\rightarrow 3}}=\frac{n^\mathrm{eq}_{1}}{R_{12\rightarrow 3}}\;,
\end{align}
where $n^\mathrm{eq}_1$ is the thermal equilibrium number density of particle $1$, and the freeze-out temperature for the given reaction $1+2\rightarrow3$ can be estimated by considering the condition $\tau_{12\rightarrow3}=1/H$.\\[-0.2cm]

\begin{figure}[ht]
\begin{center}
\hspace*{0.4cm}\includegraphics[width=0.95\linewidth]{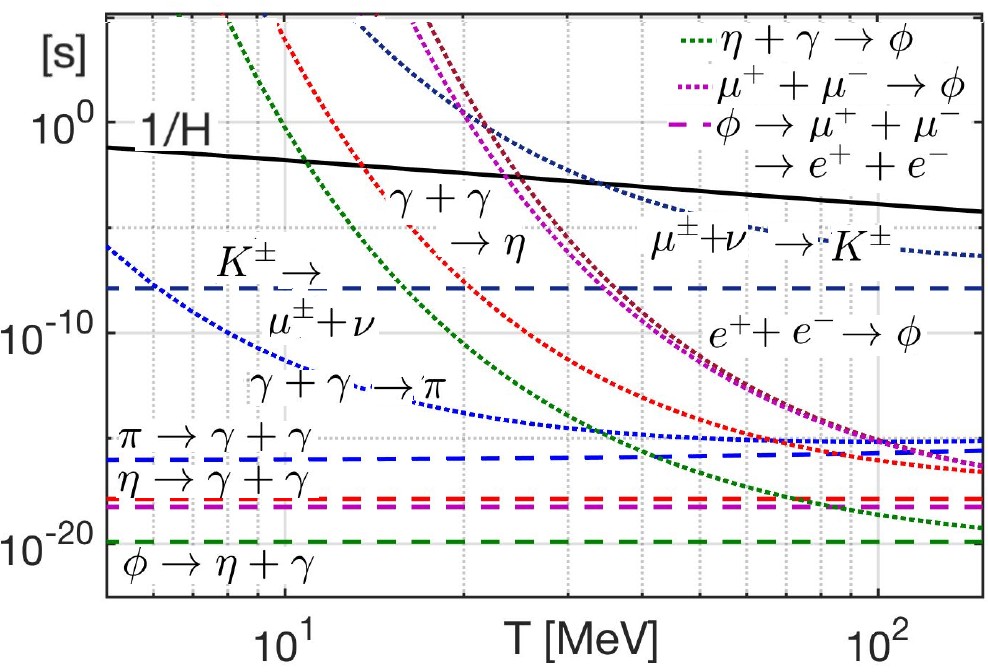}
\caption{Hubble time $1/H$ (black line) as a function of temperature is compared to leptonic and photonic relaxation reaction times, see Eq.\,(\ref{Reaction_Time}), for $\gamma+\gamma\leftrightarrow\pi$ (blue), $\gamma+\gamma\leftrightarrow\eta$ (red), $\eta+\gamma\leftrightarrow\phi$ (green), $l^++l^-\leftrightarrow\phi$ (brown and purple), and $\mu^\pm+\nu_{\mu}\longrightarrow K^\pm$(dark blue line). The horizontal dashed lines are the natural decay lifespans.}
\label{reaction_time_tot003}
\end{center}
\end{figure}

\noindent\textbf{Lepton and photon based strangeness production:} The relevant interaction rates competing with Hubble time involving lepton background in the Universe are seen in Fig.~\ref{reaction_time_tot003}. As example consider the reaction $K^\pm\leftrightarrow\mu^\pm+\nu_\mu$. For the decay modes of charged kaon $K^\pm$, we have
\begin{align}
&K^+\longrightarrow\mu^++\nu_\mu,\,\,\,\,\,\,\,\,K^-\longrightarrow\mu^-+\overline{\nu}_\mu\;,
\end{align}
where the lifetime of $K^\pm$ is given by $\tau_{K^\pm}=1.238\times10^{-8}\,\mathrm{sec}$ and the mass $m_{K^\pm}=493.677\,\mathrm{MeV}$. Hence the charged kaons in the cosmic plasma can be produced by the inverse of the decay processes, and we have
\begin{align}
&\mu^++\nu_\mu\longrightarrow K^+,\,\,\,\,\,\,\,\,\mu^-+\overline{\nu}_\mu\longrightarrow K^-\;.
\end{align}
From Eq.(\ref{Thermal_Rate}) the thermal interaction rate per unit volume for the process ${\mu^\pm+\nu\leftrightarrow K^\pm}$ can be written as
\begin{align}
R_{\mu\nu\leftrightarrow K^\pm}=\frac{g_{K^\pm}}{2\pi^2}\left(\frac{T^3}{\tau_{K^\pm}}\right)\left(\frac{m_{K^\pm}}{T}\right)^2\,K_1(m_{K^\pm}/T)\;.
\end{align}

Given the thermal interaction rate per volume $R_{\mu\nu\leftrightarrow K^\pm}$, we use for the process $\mu^\pm+\nu_{\mu}\rightarrow K^\pm$ the definition of the equilibrium relaxation time Eq.(\ref{Reaction_Time}) to obtain 
\begin{align}
&\tau_{\mu\nu\rightarrow K^\pm}=
\frac{g_\mu}{2\pi^2R_{\mu\nu\leftrightarrow K^\pm}}\int_{m_\mu}^\infty\!\!\!\!\!dE\,\frac{E\,\sqrt{E^2-m_\mu^2}}{\exp{\left(E/T\right)}+1}\;. 
\end{align} 
In Fig.(\ref{reaction_time_tot003}) the dark blue line is the interaction rate for $\mu^\pm+\nu_{\mu}\rightarrow K^\pm$.

The freeze-out condition, $\tau_{\mu\nu\rightarrow K^\pm}(T_f)=1/H(T_f)$, where the relaxation reaction time $\mu^\pm+\nu_{\mu}\rightarrow K^\pm$ becomes slower compared to the Universe expansion, and thus detailed balance in this reaction cannot be maintained, is near $T_f^{K^\pm}=33.90\,\mathrm{MeV}$. 

If other strangeness production reactions did not exist, strangeness would disappear as the Universe cools below $T_f^{K^\pm}=33.90\,\mathrm{MeV}$. However, using Eq.\,(\ref{Thermal_Rate}) and Eq.\,(\ref{Reaction_Time}) we already evaluated the relaxation time for the reactions $\pi+\pi\leftrightarrow K$, $\pi+\pi\leftrightarrow\rho$, and $\rho+\pi\leftrightarrow\phi$ as functions of temperature in the early Universe as seen in Fig. {\ref{reaction_time_tot}}. The intersection of characteristic reaction times with $1/H$ occurs for $\pi+\pi\rightarrow K$ at $T=19.85\,\mathrm{MeV}$, for $\pi+\pi\rightarrow\rho$ at $T=12.36\,\mathrm{MeV}$, and for $\rho+\pi\rightarrow\phi$ at $T=5.36\,\mathrm{MeV}$. 

Several other leptonic and photonic reactions of relevance are shown in Fig.~\ref{reaction_time_tot003}. We see now the relaxation time for $\gamma+\gamma\leftrightarrow\pi$ (blue), $\gamma+\gamma\leftrightarrow\eta$ (red), $\eta+\gamma\leftrightarrow\phi$ (green), $l^++l^-\leftrightarrow\phi$ (brown and purple), and $\mu^\pm+\nu_{\mu}\rightarrow K^\pm$(dark blue line) as a function of temperature. We see that pions remain in equilibrium since $\gamma+\gamma\rightarrow\pi$ reactions are always faster compared to $1/H$. Considering other reactions: $\gamma+\gamma\rightarrow\eta$ reaction intersects $1/H$ at $T=13.5\,\mathrm{MeV}$, $\eta+\gamma\rightarrow\phi$ at $T=10.85\,\mathrm{MeV}$; $e^-+e^+\rightarrow\phi$ at $T=24.9\,\mathrm{MeV}$; $\mu^++\mu^-\rightarrow\phi$ at $T=23.5\,\mathrm{MeV}$; and as noted before, $\mu^\pm+\nu_{\mu}\rightarrow K^\pm$ at $T=33.9\,\mathrm{MeV}$.\\[-0.2cm]

\noindent\textbf{Discussion:}
In this work we have explored reaction rates in the meson sector of strangeness. We have identified hierarchy of reactions which become slower compared to Hubble time $1/H$ involving strange particles $\phi$ and $K$. The reactions $\mu^\pm+\nu_{\mu}\rightarrow K^\pm$, $l^-+l^+\rightarrow\phi$, and $\pi+\pi\rightarrow K$ are intersecting $1/H$ at $T=33.9\,\mathrm{MeV}$, $T=25\,\mathrm{MeV}$, and $T=20\,\mathrm{MeV}$, respectively. 

Once these (or any other) reactions decouple from the cosmic plasma, the corresponding detailed balance is broken: Absence of the inverse decay reaction acts like a ``hole'' in the strangeness abundance ``pot'' with some strangeness decay in early Universe proceeding. However, the corresponding reverse channel of strangeness production reaction is too slow compared to the fast Universe expansion. Considering only meson content of the Universe, our results show that when the expanding Universe cools to a temperature around $T=20\,\mathrm{MeV}$, strangeness will begin to progressively decouple but remain near-equilibrium abundance.

However, there are many different channels of strangeness production and decay so loss of one reaction channel does not end the strangeness abundance. As the Universe expands, other strangeness producing reactions become relevant; for example, $\gamma+\gamma\rightarrow\eta$ reaction intersects $1/H$ at $T=13.5\,\mathrm{MeV}$, and $\eta+\gamma\rightarrow\phi$ at $T=10.85$\,MeV. Thus when an $\eta$ is produced, we can always assume that a $\phi$ arises and thus considering the $\phi\leftrightarrow K+K$ reactions this generates Kaons which can decay as now several strangeness decay paths exceed in speed the corresponding channel production rate. We conclude that strangness chemical decoupling is complete once all dominant strangness formation reactions become too slow.

We have considered kinetically equilibrated distribution described by the temperature $T$. This certainly applies to charged particles which are embedded in the cosmic background of a dense photon-electron-positron plasma. However, many of the reactions we considered involve neutral particles; as examples consider the process $\pi^0+\pi^0\to K_\mathrm{S}$, and many even more relevant reactions involving neutral strange baryons. Hadrons have relatively large strong interaction reaction cross sections. However, as the temperature decreases, their abundance is very small. Further study needs to explore the validity of kinetic equilibrium hypothesis for neutral hadrons.

This first study of strangeness production mechanisms in the early Universe, following on QGP hadronization, shows that the relevant temperature domains indicating non-trivial dynamics and freeze-out for (strange) mesons and strange (anti)baryon abundance are practically overlapping. Thus both baryon and strangeness chemical freeze-out have to be explored jointly in a dynamical model. The Universe in the range $20\le T\le 60\,\mathrm{MeV}$ appears rich in physics phenomena involving strange mesons, (anti)baryons including (anti)hyperon abundances, which can enter into chemical nonequilibrium while undergoing successive freeze-out processes. This means that a quantitative study of strangeness freeze-out requires that a full account be given to strange hyperons and antihyperons. For further discussion we refer to our forthcoming work~\cite{Yang:2021bko}.\\[-0.2cm]

\noindent\textbf{Acknowledgment:} The research of CTY was supported by the US Department of Energy under Grant Contract DESC0012704 to the Brookhaven National Laboratory. We thank J. Birrell for interesting discussions.



\begin{thebibliography}{99}
 
\bibitem{Letessier:2002gp} 
J.~Letessier and J.~Rafelski,
\emph{Hadrons and Quark-Gluon Plasma}, 
397 pp, Camb.\ Monogr.\ Part.\ Phys.\ Nucl.\ Phys.\ Cosmol.\ {\bf 18} (2002)
doi: 10.1017/CBO9780511534997,
ISBN: 9780521018234 (Paperback), 9780521385367 (Hardback), 9780511037276 (Online)

\bibitem{Kolb:1990vq} 
E.~W.~Kolb and M.~S.~Turner,
\emph{The Early Universe},
547 pp, Front.\ Phys.\ {\bf 69}, 1 (1990),
ISBN: 0201626748, 9780201626742

\bibitem{Rafelski:2021aey}
J.~Rafelski and C.~T.~Yang,
``The muon abundance in the primordial Universe,''
\doi{10.5506/APhysPolB.52.277}{Acta Phys. Polon. B \noindent\textbf{52} (2021), 277}
[arXiv:2103.07812 [hep-ph]].

\bibitem{Kuznetsova:2008jt} 
 I.~Kuznetsova, D.~Habs and J.~Rafelski,
 ``Pion and muon production in e-, e+, gamma plasma,''
 \doi{10.1103/PhysRevD.78.014027}{Phys.\ Rev.\ D {\bf 78}, 014027 (2008)}
 [arXiv:0803.1588 [hep-ph]].
 
\bibitem{Birrell:2014uka}
J.~Birrell, C.~T.~Yang and J.~Rafelski,
``Relic Neutrino Freeze-out: Dependence on Natural Constants,''
\doi{10.1016/j.nuclphysb.2014.11.020}{Nucl. Phys. B \noindent\textbf{890}, 481-517 (2014)}
[arXiv:1406.1759 [nucl-th]].

\bibitem{Tanabashi:2018oca} 
M.~Tanabashi {\it et al.} [Particle Data Group],
``Review of Particle Physics,''
Phys.\ Rev.\ D {\bf 98}, no. 3, 030001 (2018)
doi:10.1103/PhysRevD.98.030001

\bibitem{Koch:1986ud}
P.~Koch, B.~Muller and J.~Rafelski,
``Strangeness in Relativistic Heavy Ion Collisions,''
Phys. Rept. \noindent\textbf{142}, 167-262 (1986)
doi:10.1016/0370-1573(86)90096-7

\bibitem{Yang:2021bko}
C.~T.~Yang and J.~Rafelski,
``Cosmological Strangeness Abundance,''
[arXiv:2108.01752 [hep-ph]].
\end{thebibliography}
\end{document}